\documentclass[journal]{IEEEtran}
\IEEEoverridecommandlockouts
\usepackage{cite}
\usepackage{amsmath,amssymb,amsfonts}
\usepackage{graphicx}
\usepackage{textcomp}
\usepackage{xcolor}

\usepackage{hyperref}
\usepackage{amsmath}
\usepackage{autobreak}
\usepackage{amsfonts}
\usepackage{bm}
\usepackage{algorithm}  
\usepackage{algorithmicx}  
\usepackage{algpseudocode}
\usepackage{stfloats}
\usepackage{cuted}
\usepackage{color}
\usepackage{subfigure}
\definecolor{red}{RGB}{0,0,0}

\newcommand{\st}{\mathbf{{\rm s.t.}}}

\def\BibTeX{{\rm B\kern-.05em{\sc i\kern-.025em b}\kern-.08em T\kern-.1667em\lower.7ex\hbox{E}\kern-.125emX}}
\makeatletter
\renewcommand{\maketag@@@}[1]{\hbox{\m@th\normalsize\normalfont#1}}% 
\makeatother % 公式字体变小但编号大小不变

\begin{document}

\title{Age Minimization in Outdoor and Indoor Communications with Relay-aided Dual RIS
}

\author{Wanting Lyu,~Yue Xiu,~Yang Zhao,~Chadi Assi,~\IEEEmembership{Fellow,~IEEE}~and~Zhongpei Zhang,~\IEEEmembership{Member,~IEEE} \\
	
	\thanks{Wanting Lyu, Yue Xiu and Zhongpei Zhang are with National Key Laboratory of Science and Technology on Communications, University of Electronic Science and Technology of China, Chengdu61173, China (E-mail: lyuwanting@yeah.net; xiuyue12345678@163.com; zhangzp@uestc.edu.cn).
	
	Yang Zhao is with Singapore Institute of Manufacturing Technology, A*STAR, Singapore, 138634 (Email: zhao\_yang@simtech.a-star.edu.sg).
		
	Chadi Assi is with the Concordia Institute for Information
    Systems Engineering, Concordia University, Montreal, QC H3G 1M8, Canada (e-mail: assi@ciise.concordia.ca).
	}
}

\maketitle

\begin{abstract}
In this paper, we investigate an outdoor and indoor wireless communication network with the assistance of a relay-aided double-sided reconfigurable intelligent surface (RIS). A scheduling problem is considered at the outdoor access point (AP) to minimize the sum of age of information (AoI). To serve the indoor users and further enhance the wireless link quality, a double-sided RIS with relay is utilized. Since the formulated problem is non-convex with highly-coupled variables, a successive convex approximation (SCA) and penalty based alternating optimization (AO) algorithm with difference of convex (DC) functions is proposed to solve it in an iterative manner. Finally, simulation results show the effectiveness and significant performance improvement in terms of AoI of the proposed algorithm compared with other baselines.
\end{abstract}

\vspace*{-0.15cm}
\section{Introduction}
% Existing cellular networks are facing significant challenges resulting from increasing demands for better link quality and coverage range. 
The target of the fifth generation (5G) and beyond mobile communication system is to cater services for emerging applications such as ultra-reliable low latency communications (URLLC), enhanced mobile broadband (eMBB) and massive machine-type communications (mMTC) \cite{123456}. More recently, some emerging applications (such as industrial control and sensing, cooperative autonomous driving (CAD), etc.) have accelerated the demand for information freshness and real-time status updates \cite{LBHuang7282742}; for example, a CAD application relies on a multitude of sensory data with rigorous requirement on timeliness, such as vehicles speed, precise location, direction, etc. \cite{6603997}. To quantify the freshness of information, a new indicator has emerged, namely the age of information (AoI) \cite{AoI6195689}, which is defined as the time that has elapsed since the generation of the last successfully delivered status update packet. Existing work tackled the optimization of the AoI in different application scenarios, such as wireless powered networks \cite{AoI_wireless_powered} and multi-access edge computing (MEC)-assisted Internet of things (IoT) networks \cite{AoIMEC}. 

Indeed, a key requirement for enabling the above services is to have a network in place with the corresponding enabling technologies and a variety of such technologies have emerged as enablers for 5G and future networks; for instance, non orthogonal multiple access (NOMA), massive multiple-input multiple-output (MIMO), small cells, cloud radio access network (RAN), cellular connected unmanned aerial vehicles (UAVs), etc. do promise to enhance spectral efficiency, increase connectivity, expand coverage and improve the quality of communication links and hence data rates. More recently, reconfigurable intelligent surface (RIS) has emerged and is considered as a promising technology to alleviate the severe attenuation in wireless propagation \cite{Marco9481223}.  Interestingly, RIS has been integrated into real-time networks to optimize the freshness of information \cite{Ali,LWT,LWTaerial,AoI_RIS_NOMA}. 

However, traditional RIS requires the users to be located at the same side of the RIS with only half-space coverage, and cannot serve the outdoor and indoor users simultaneously. To overcome this drawback, the authors in \cite{STARris360, STARris} proposed a novel simultaneously transmitting and reflecting (STAR)-RIS, extending the coverage range from half-space to full-space. However, the passive STAR-RIS still has limitations on the quality of service (QoS) due to the severe attenuation caused by the double fading effect of RIS. To enhance the communication quality, the authors in \cite{RelayaidedIRS} proposed a novel relay-aided RIS architecture, where a full-duplex relay was used to amplify the signal between the two RISs. This model significantly reduces the number of reflecting elements compared with the passive RIS. Following this, in \cite{dual_function}, a dual relay and reflection RIS was proposed to maximize the achievable sum rate, showing remarkable performance improvement compared with full-duplex relay and STAR-RIS. {\color{red} However, the above works only considered the optimization of data rate, which is far from enough for the rapidly evolving real-time applications.}

{\color{red} To overcome this}, we study a real-time outdoor and indoor multi-user wireless network with a relay-aided double-sided RIS. {\color{red} To maintain the information freshness while guaranteeing the data rate,} a joint user scheduling, active beamforming and phase shifts design problem is formulated to minimize the sum AoI. To tackle the non-convexity with highly coupled variables, we propose an SCA based AO algorithm with difference of convex (DC) method to decompose the intractable problem into two sub-problems and solve them iteratively. Numerical simulations show the considerable performance improvement of the proposed algorithm based on relay-aided RIS compared with baselines including full duplex amplify-and-forward (AF) relay, random phase shift, and random beamforming scheme.

\section{System Model and Problem Formulation}

\begin{figure}[t]
    \centering
    \includegraphics[width=0.7\linewidth]{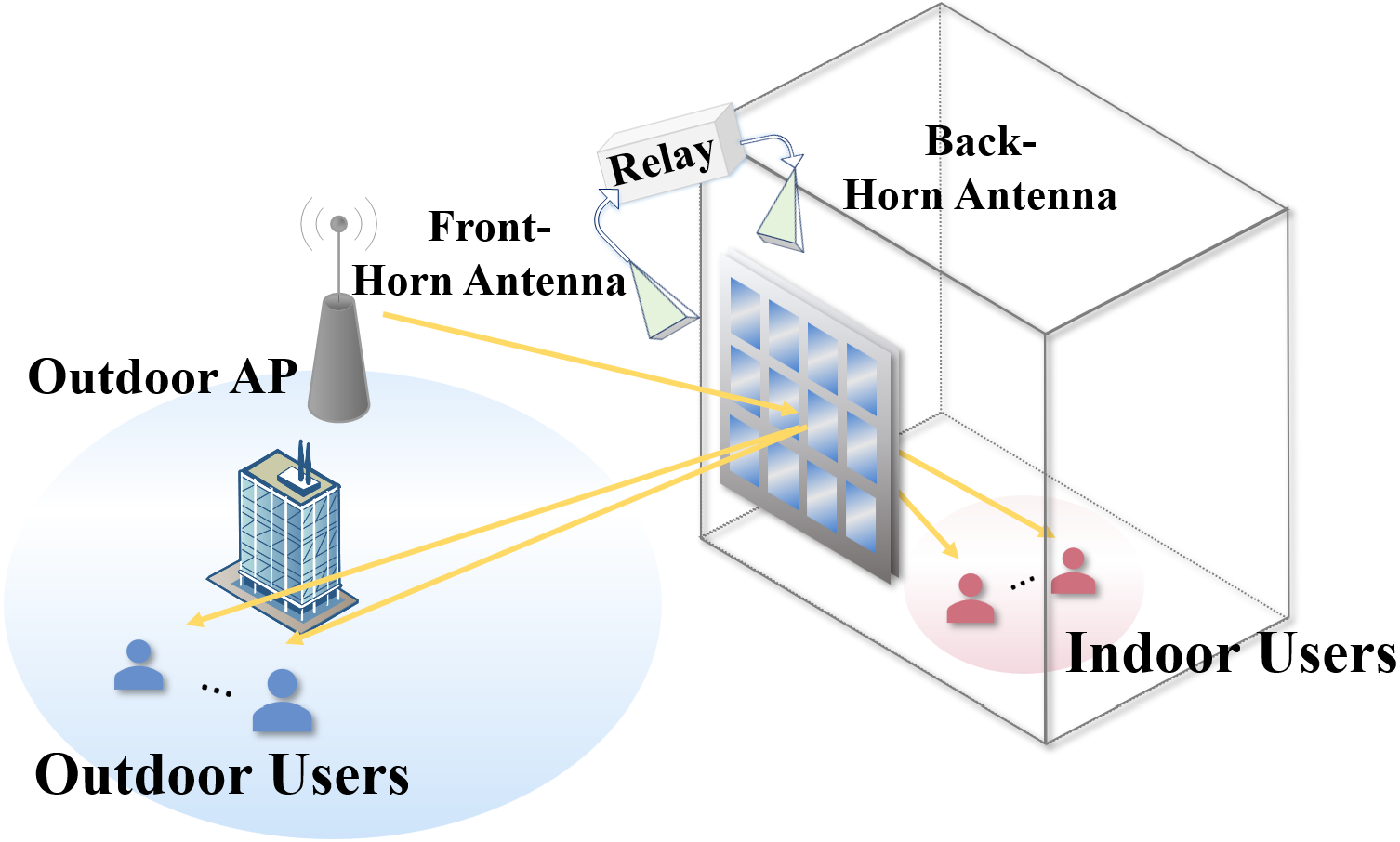}
    \caption{System model for the wireless network.}
    \label{model}
\end{figure}

Consider a downlink wireless network with a double-sided RIS assisted by an AF relay deployed on the building wall as illustrated in Fig. \ref{model}. Assume that the outdoor access point (AP) is equipped with $M$ antennas serving $J$ outdoor and $K$ indoor single-antenna users that are located in front of and at the back of the RIS. At the AP, there are $J+K$ data streams with randomly arriving packets corresponding to each user. The direct links between the AP and the users are assumed to be blocked. Consider a reflecting and amplifying RIS model, where the front surface and the back surface are both equipped with $N_s$ reflecting elements. For the indoor users, the incident signal from the AP is reflected first by the front RIS to the front horn antenna, and then amplified by the relay. Before being reflected to the indoor users, the signal is emitted to the back-RIS by the back-horn antenna.

The time axis is divided into time slots, lasting for $T$ slots in total. The spectral resources for different users are assumed to be orthogonal so that the mutual interference is avoided. The channel between the AP and the front-RIS is denoted as $\mathbf{G}(t) \in \mathbb{C}^{N_s\times M}$, while the channel from the front-RIS to the outdoor front-user $j$ and the channel from the back-RIS to the indoor back-user $k$ are $\big(\mathbf{h}_{rj}^F(t)\big)^H$ and $\big(\mathbf{h}_{rk}^{B}(t) \big)^H\in \mathbb{C}^{1\times N_s}$, respectively. The channels between the RIS and front- and back- horn antennas are $\mathbf{g}_f^H(t) \in \mathbb{C}^{1\times N_s} $ and $\mathbf{g}_b(t) \in \mathbb{C}^{N_s\times 1}$, respectively. We define phase shift matrices $\Phi_f(t) = {\rm diag}\Big(e^{j\theta_1^F(t)}, e^{j\theta_2^F(t)},..., e^{j\theta_{N_s}^F(t)}\Big)$ for the front-RIS, and $\Phi_b(t) = {\rm diag}\Big(e^{j\theta_1^B(t)}, e^{j\theta_2^B(t)},..., e^{j\theta_{N_s}^B(t)}\Big)$ for the back-RIS, where $\theta_n^F(t), \theta_n^B(t) \in [0, 2\pi), \forall n\in \{1,...,N_s\}$.

At each time slot, the users are scheduled at the AP with the indicator $a_j^F(t) \in \{0,1\}$, where $a_j^F(t)=1$ means stream $j$ is scheduled to transmit a packet to front-user $j$ at slot $t$, and those for back-users are defined similarly. The number of total available channels is assumed to be $E$, which can be expressed as
% \begin{small}
\begin{equation}
    \sum_{j=1}^J a_j^F(t)+\sum_{k=1}^K a_k^B(t) \le E.
    \label{Cons_avai_chann}
\end{equation}
% \end{small}%
The transmit symbols at the AP are denoted as $s_j^F,\,s_k^B(t) \sim \mathcal{CN}(0,1)$. The transmit beamforming vectors for front-user $j$ and back-user $k$ are $\mathbf w_j^F,\text{ and }\mathbf{w}_k^B(t) \in \mathbb{C}^{M\times 1}$. The transmit power budget is $P_0$, and thus the active beamforming vectors are constrained as 
\begin{equation}
    \sum_{j=1}^{J}||\mathbf{w}_j^F(t)||^2 + \sum_{k=1}^{K}||\mathbf{w}_k^B(t)||^2 \le P_0.
    \label{Cons_pow}
\end{equation}

Hence, the received signal of back-user $k$ at the input of the relay is denoted as 
\begin{equation}
    y_{fk}^B(t) = {\color{red}\mathbf{g}_{f}^H(t)}\mathbf{\Phi}_f(t)\mathbf{G}(t)\mathbf{w}_k^B(t)s_k^B(t) + n_o(t),
    \label{Rx_amp}
\end{equation}
\noindent where the thermal noise is denoted as $n_o(t) \sim \mathcal{CN}(0,\sigma_o^2)$. Thus, the received signal of back-user $k$ can be expressed as 
\begin{small}
\begin{align}
    y_k^B(t) = \sqrt{\chi}\big(\mathbf{h}_{rk}^B(t)\big)^H\mathbf{\Phi}_b(t)\mathbf{g}_b(t) \Big( \mathbf{g}_f^H(t)\mathbf{\Phi}_f(t)\mathbf{G}(t) \nonumber \\
    \mathbf{w}_k^B(t)s_k^B(t) 
     + n_o(t) \Big) + n_k^B(t),
    \label{Rx_user}
\end{align}
\end{small}
\noindent where the amplification gain $\chi$ is assumed to be a known constant, and the additive white Gaussian noise (AWGN) is denoted as $n_k^B(t)  \sim \mathcal{CN}(0,\sigma_k^{B2})$.

Accordingly, the received SNR of back-user $k$ can be expressed as
\begin{small}
\begin{equation}
    \gamma_k^B(t) = \frac{ \chi \Big|\big(\mathbf{h}_{rk}^B(t)\big)^H\mathbf{\Phi}_b(t) \mathbf{g}_b(t)\mathbf{g}_f^H(t)\mathbf{\Phi}_f(t)\mathbf{G}(t)\mathbf{w}_k^B(t)\Big|^2 } {\chi\Big|\big(\mathbf{h}_{rk}^B(t)\big)^H\mathbf{\Phi}_b(t)\mathbf{g}_b(t)\Big|^2\sigma_o^2+ \sigma_k^{B2}}.
    \label{SNR_Buser}
\end{equation}
\end{small}
\noindent The received SNR of front-user $k$ is the same as the traditional RIS, which can be written as
\begin{small}
\begin{equation}
    \gamma_j^F(t) = \frac{\Big|\big(\mathbf{h}_{rj}^F(t)\big)^H\mathbf \Phi_f(t)\mathbf G(t)\mathbf w_j^F(t)\Big|^2 }{\sigma_j^{F2}}.
    \label{SNR_Fuser}
\end{equation}
\end{small}

\noindent A successful delivery of a packet requires the received SNR to be greater than or equal to the SNR threshold $\gamma_{th}$, namely
\begin{small}
\begin{equation}
    \gamma_j^F(t) \ge \gamma_{th},\;\gamma_k^B(t) \ge \gamma_{th}.
    \label{SNR_thresh}
\end{equation}
\end{small}%

In this system, we introduce the concept of the AoI to quantify the freshness of information. At the AP, we design the scheduling policy to minimize the sum AoI. For simplicity and without loss of generality, we take front-user $j$ as an example to elaborate the definitions for the parameters, where those for back-users are defined in the same manner.

The packets arrive randomly in queue at the AP, where only the latest packet of each data stream can be stored. Define a packet arrival indicator $p_j^F(t) \in \{0,1\}$ with $p_j^F(t) = 1$ suggesting that a new packet of data stream $j$ arrives at the AP at the beginning of time slot $t$. The system time $z_j^F(t)$ is defined as in \cite{Ali}, which will be set as $0$ when a new packet arrives at the queue, and otherwise it will increase linearly with time. Hence, it can be expressed as
\begin{equation}
z_j^F(t) = \left\{
\begin{aligned}
&0,  &{\rm if}\; p_j^F(t) = 1,  \\
&z_j^F(t-1)+1 , &{\rm otherwise}.
\end{aligned}
\right.
\label{sys_time}
\end{equation}

\noindent The instantaneous AoI is represented as $A_j^F(t), \; j\in\{1,...,J\}$, whose value will be updated if a new status update data packet is successfully received by the user, and otherwise it will increase linearly. Indicator $\eta_j^F(t) \in \{0,1\}$ suggests whether stream $j$ has an available packet to be delivered, which is expressed as in \cite{Ali},
\begin{equation}
    \eta_j^F(t) = p_j^F(t) + \eta_j^F(t-1)(1-a_j^F(t-1))(1-p_j^F(t)).
\end{equation}

\noindent Accordingly, $A_j^F(t)$ can be expressed as 
\begin{small}
\begin{equation}
A_j^F(t)=\left\{
\begin{aligned}
& z_j^F(t-1) + 1,  & {\rm if}\; a_j^F(t-1)\eta_j^F(t-1) = 1 \; \\
&   & {\rm and} \; \gamma_j^F(t-1) \ge \gamma_{th}, \\ 
& A_j^F(t-1) + 1, & {\rm otherwise},
\end{aligned}
\right.
\end{equation}
\end{small}

\begin{figure}[!t]
    \centering
    \includegraphics[width=0.7\linewidth]{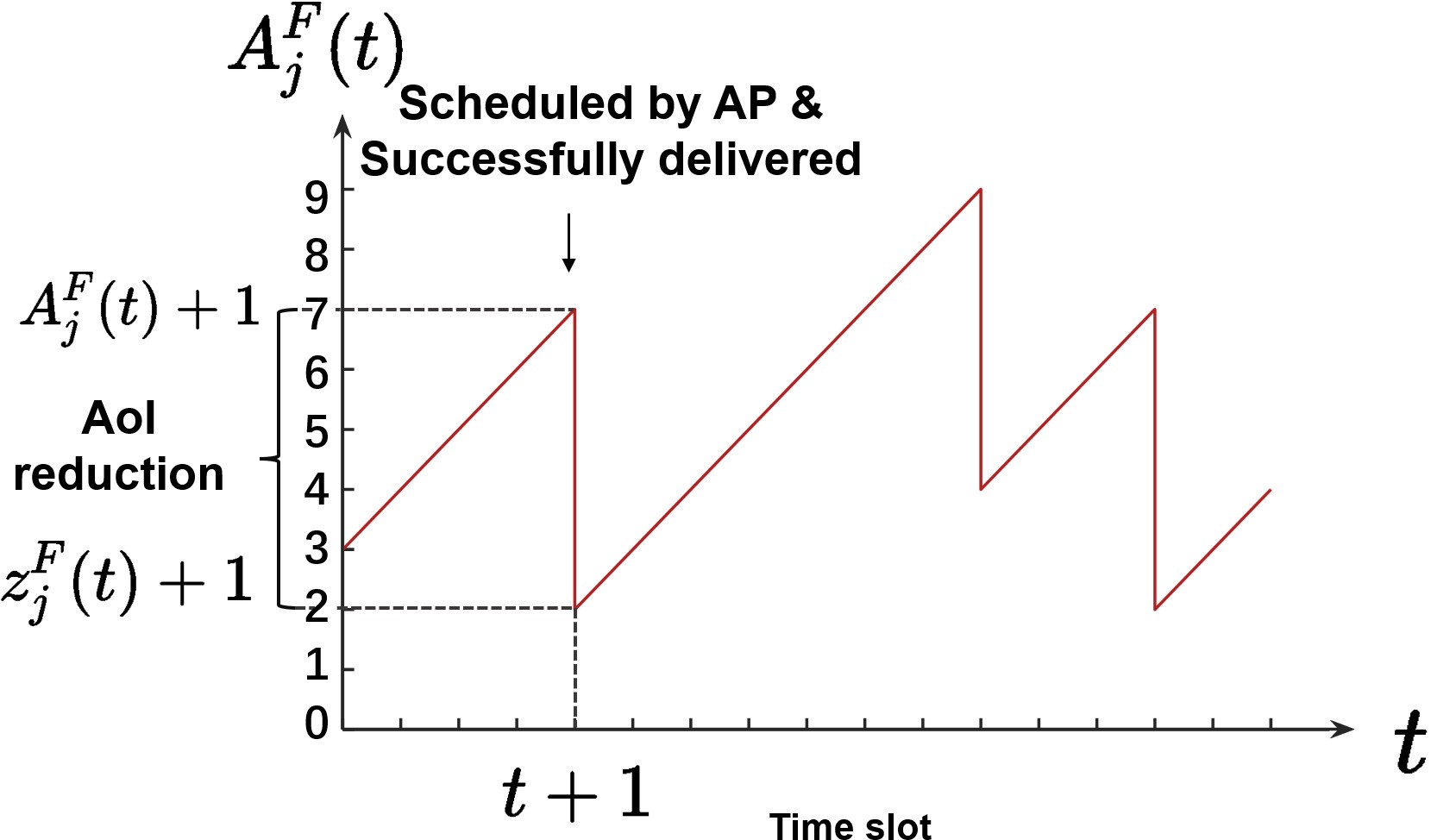}
    \caption{AoI evolution along time slot.}
    \label{AoI_evolution}
\end{figure}

\noindent Fig. \ref{AoI_evolution} illustrates the AoI evolution along the time slot, where the stream is scheduled by the AP at the beginning of time slot $t$ and successfully received and decoded by the user at $t+1$. Thus, the corresponding AoI drops from $A_j^F(t)+1$ to $z_j^F(t)+1$, and the amount of AoI reduction is hence $A_j^F(t) - z_j^F(t)$.

To guarantee the information freshness of the system, the problem is formulated to minimize the sum AoI by jointly optimizing the user scheduling with active and passive beamforming design. Therefore, the corresponding optimization problem can be expressed as
\begin{small}
\begin{align}
    \mathrm{(P1)}:& \min_{a_j^F(t),a_k^B(t),\mathbf w_j^F(t),\atop \mathbf w_k^B(t), \mathbf \Phi_f(t), \mathbf \Phi_b(t)}  \sum_{t=1}^T\Big ( \sum_{j=1}^J A_j^F(t) + \sum_{k=1}^K A_k^B(t)  \Big)
    \label{P1obj}\\
    \st \quad & (\ref{Cons_avai_chann}),(\ref{Cons_pow}), \nonumber \\
    & a_j^F(t) \in \{0,1\}, \; a_k^B(t) \in \{0,1\},
    \label{Cons_sche}\tag{\ref{P1obj}{a}} \\
    & \gamma_j^F(t) \ge \gamma_{th}a_j^F(t)\eta_j^F(t),
    \label{Cons_FSNR}\tag{\ref{P1obj}{b}} \\
    & \gamma_k^B(t) \ge \gamma_{th}a_k^B(t)\eta_k^B(t),
    \label{Cons_BSNR}\tag{\ref{P1obj}{c}} \\
    & \theta_n^F \in [0,2\pi),\; \theta_n^B \in [0,2\pi).
    \label{Cons_theta} \tag{\ref{P1obj}{d}} 
\end{align}
\end{small}%
Note that the AoI will keep increasing linearly without newly decoded status update packet, resulting in undesired sum AoI throughout the given $T$ time slots. Therefore, to minimize the sum AoI in the total period, it is effective to maximize the sum AoI reduction in each time slot as shown in Fig. \ref{AoI_evolution} \cite{SHZhang9130055}. Hence, the problem can be reformulated as (P2) at the bottom of this page. To solve the highly-coupled non-convex problem, we decouple the variables by AO algorithm and DC functions, and iteratively tackle the non-convexity by SCA algorithm in section III.
\begin{figure*}[b]
\hrule
\begin{small}
\begin{align}
    \mathrm{(P2)}: \max_{a_j^F(t),a_k^B(t),\mathbf w_j^F(t),\atop \mathbf w_k^B(t), \mathbf \Phi_f(t), \mathbf \Phi_b(t)} \mathcal{G}(a_j^F(t), a_k^B(t)) = &
    \sum_{j=1}^J \Big(A_j^F(t) - z_j^F(t)\Big)a_j^F(t)\eta_j^F(t) + \sum_{k=1}^K \Big(A_k^B(t) - z_k^B(t)\Big)a_k^B(t)\eta_k^B(t) 
    \label{P2obj}\\
    \st \quad & (\ref{Cons_avai_chann}),(\ref{Cons_pow}),(\ref{Cons_sche}),(\ref{Cons_FSNR}),(\ref{Cons_BSNR}),(\ref{Cons_theta}). \nonumber
\end{align}
\end{small}
\begin{small}
\begin{gather}
    \gamma_j^F(t) = \frac{|\mathbf\psi_f^H(t){\rm diag}\{\big(\mathbf{h}_{rj}^F(t)\big)^H\}\mathbf G(t)\mathbf w_j^F(t)|^2}{\sigma_j^{F2}} \ge \gamma_{th}a_j^F(t)\eta_j^F(t), 
    \label{Cons_SNR_Fupd}
    \tag{14}\\
    \gamma_k^B(t)=\frac{ \chi|\mathbf{\psi}_b^H(t){\rm diag}\{\big(\mathbf h_{rk}^{B}(t)\big)^H\}\mathbf g_b(t)  \mathbf{\psi}_f^H(t) {\rm diag}\{\mathbf g_f^H(t)\} \mathbf G(t)\mathbf w_k^B(t)  |^2 }{ \chi|\mathbf \psi_b^H(t){\rm diag}\{\big(\mathbf h_{rk}^{B}(t)\big)^H\}\mathbf g_b(t)|^2\sigma_o^2 + \sigma_k^{B2} } \ge \gamma_{th}a_k^B(t)\eta_k^B(t),
    \label{Cons_SNR_Bupd}
    \tag{15}
\end{gather}
\end{small}
\end{figure*}
% 底部的（15）

% \begin{figure*}[b]
% \begin{equation}
%     \gamma_k^B(t)=\frac{ \chi|\mathbf{\psi}_b^H(t){\rm diag}\{\mathbf h_{rk}^{BH}(t)\}\mathbf g_b(t) {\rm diag}\{\mathbf g_f^H(t)\} \mathbf G(t)\mathbf w_k^B(t) \mathbf{\psi}_f^H(t) |^2 }{ |\mathbf \psi_b^H(t){\rm diag}\{\mathbf h_{rk}^{BH}(t)\}\mathbf g_b(t)|^2\sigma_o^2 + \sigma_k^{B2} } \ge \gamma_{th}a_k^B(t)\eta_k^B(t),
%     \label{SNR_Bupd}
%     \tag{15}
% \end{equation}
% \end{figure*}

\section{Proposed Solutions}
In this section, the proposed AO algorithm is elaborated as follows. First, the channel is rewritten in a more tractable form by defining vectors $\mathbf\psi_f(t)$ and $\mathbf\psi_b(t)$, with $\mathbf{\psi}_f^{\rm H}(t) = [e^{j\theta_1^F(t)}, ..., e^{j\theta_{N_s}^F(t)}] \in \mathbb{C}^{1\times N_s}$ and $\mathbf{\psi}_b^{\rm H}(t) = [e^{j\theta_1^B(t)}, ..., e^{j\theta_{N_s}^B(t)}] \in \mathbb{C}^{1\times N_s}$, respectively. This results in unit modulus constraints for each element in $\mathbf\psi_f^H(t)$ and $\mathbf\psi_b^H(t)$, namely
\begin{equation}
    |\psi_{fn}(t)| = 1, \;|\psi_{bn}(t)| = 1.
    \label{Cons_modu}
\end{equation}
In this way, the received SNR constraints can be reformulated as (\ref{Cons_SNR_Fupd}) and (\ref{Cons_SNR_Bupd}) at the bottom.

Since the active and passive beamforming vectors are strongly coupled, the formulated problem is decomposed into two sub-problems to be alternatively solved.
\vspace{-0.4cm}
\subsection{Active Beamforming and User Scheduling Design}

Given $\mathbf \psi_f(t)$ and $\mathbf \psi_b(t)$, the problem becomes
\begin{small}
\begin{align}
    \mathrm{(P3.1)} & \max_{a_j^F(t),a_k^B(t),\atop \mathbf w_j^F(t), \mathbf w_k^B(t)}  
    \mathcal{G}(a_j^F(t), a_k^B(t))
    \label{P3.1obj}     \tag{16} \\
    \st \quad & (\ref{Cons_avai_chann}),(\ref{Cons_pow}),(\ref{Cons_sche}),(\ref{Cons_SNR_Fupd}),(\ref{Cons_SNR_Bupd}). \nonumber
\end{align}
\end{small}

\noindent First, the binary constraint (\ref{Cons_sche}) is non-convex and is relaxed as \cite{scheduling}
\begin{small}
\begin{equation}
    a_j^F(t) \in [0,1],\; a_k^B(t) \in [0,1].
    \label{Cons_sche_relax}\tag{17}
    \vspace{-0.2cm}
\end{equation}
\end{small}
\noindent To deal with the non-convex constraints (\ref{Cons_SNR_Fupd}), we define the equivalent cascaded channel $\big(\bar {\mathbf h}_{rj}^F(t)\big)^H = \mathbf\psi_f^H(t){\rm diag}\{\big(\mathbf{h}_{rj}^F(t)\big)^H\}\mathbf G(t)$, and put the denominator to the right side of the inequality, such that
\begin{equation}
    \Big|\big(\bar {\mathbf h}_{rj}^F(t)\big)^H \mathbf w_j^F(t)\Big|^2   \ge \gamma_{th}a_j^F(t)\eta_j^F(t)\sigma_j^{F2}.
    \label{ineq}\tag{18}
\end{equation}
\noindent The lower bound of the left hand side (LHS) of the inequality (\ref{ineq}) is approximated by its first Taylor expansion, given as
\begin{small}
\begin{align}
     & 2\mathcal{R} \Big\{ \Big( \big(\bar {\mathbf h}_{rj}^F(t)\big)^H \big(\mathbf w_j^F(t)\big)^{(i-1)}\Big)^H \big(\bar {\mathbf h}_{rj}^F(t)\big)^H \Big(\mathbf w_j^F(t) \nonumber\\ 
     & - \big( \mathbf w_j^F(t)\big)^{(i-1)} \Big) \Big\} + \Big|\big(\bar {\mathbf h}_{rj}^F(t)\big)^H\big(\mathbf w_j^F(t)\big)^{(i-1)}\Big|^2 \nonumber \\
     &\ge\; \gamma_{th}a_j^F(t)\eta_j^F(t)\sigma_j^{F2},
    \label{Cons_FSNR_SCA}\tag{19}
\end{align}
\end{small}%
\noindent where the superscript $(i-1)$ means the value in the $(i-1)^{th}$ iteration.

Similarly, for the back-user $k$, defining the equivalent cascaded channel $\big(\bar {\mathbf h}_{rk}^B(t)\big)^H = \mathbf{\psi}_b^H(t){\rm diag}\{\big(\mathbf h_{rk}^{B}(t)\big)^H\}\mathbf g_b(t)  \mathbf{\psi}_f^H(t) {\rm diag}\{\mathbf g_f^H(t)\} \mathbf G(t)$, (\ref{Cons_SNR_Bupd}) becomes (\ref{Cons_BSNR_SCA}) at the bottom of the next page.
\begin{figure*}[b]
\hrule
\begin{small}
\begin{align}
    2\mathcal{R} \Big\{ \Big( \big(\bar {\mathbf h}_{rk}^B(t)\big)^H& \big(\mathbf w_k^B(t)\big)^{(i-1)}\Big)^H \big(\bar {\mathbf h}_{rk}^B(t)\big)^H \Big( \mathbf w_k^B(t) - \big(\mathbf w_k^B(t)\big)^{(i-1)}  \Big) \Big\} 
    +\;\left|\big(\bar {\mathbf h}_{rk}^B(t)\big)^H\big(\mathbf w_k^B(t)\big)^{(i-1)}\right|^2 \nonumber  \\
    & \ge\;  \gamma_{th}a_k^B(t)\eta_k^B(t)\Big(\chi|\mathbf \psi_b^H(t){\rm diag}\{\big(\mathbf h_{rk}^{B}(t)\big)^H\}\mathbf g_b(t)|^2\sigma_o^2 + \sigma_k^{B2}\Big)
    \label{Cons_BSNR_SCA}\tag{20}
\end{align}
\begin{align}
    \Delta_{\sigma_k}(t) \le \frac{1}{4}\Bigg[\bigg({\rm Tr}\Big( \mathbf\Psi_b(t) &\mathbf H_k^B(t)\Big) + a_k^B(t) \bigg)^2 -
    2\bigg({\rm Tr}\Big( \mathbf\Psi_b^{(i-1)}(t) \mathbf H_k^B(t) \Big) - a_k^{B(i-1)} (t) \bigg)\bigg( {\rm Tr}\Big( \mathbf\Psi_b(t) \mathbf H_k^B(t) \Big) - a_k^B (t) \bigg) \nonumber \\
    + &\bigg({\rm Tr}\Big( \mathbf\Psi_b^{(i-1)}(t) \mathbf H_k^B(t) \Big) - a_k^{B(i-1)} (t) \bigg)^2\Bigg]\chi\sigma_o^2\gamma_{th}a_k^B(t)\eta_k^B(t) 
    = \Big[\Delta_{\sigma_k}(t)\Big]_{ub}
    \label{Tr_term1_ub}\tag{27} 
\end{align}
\end{small}
\end{figure*}

Therefore, in the $(i)^{th}$ iteration, problem (P3.1) becomes
\begin{small}
\begin{align}
    \mathrm{(P3.2)}& \max_{a_j^F(t),a_k^B(t),\atop \mathbf w_j^F(t), \mathbf w_k^B(t)}  \mathcal{G}(a_j^F(t), a_k^B(t))
    \label{P3.2obj}\tag{21}\\
    \st \quad & (\ref{Cons_avai_chann}),(\ref{Cons_pow}),(\ref{Cons_sche_relax}),(\ref{Cons_FSNR_SCA}),(\ref{Cons_BSNR_SCA}), \nonumber
\end{align}
\end{small}%
\noindent which is a convex problem that can be efficiently solved by standard solver such as CVX.

\subsection{RIS Phase Shifts Design and User Scheduling Updating}

Given the active beamforming vectors, problem (P2) becomes
\begin{small}
\begin{align}
    \mathrm{(P4.1)}& \max_{a_j^F(t),a_k^B(t), \atop \mathbf \psi_f(t), \mathbf \psi_b(t)}  \mathcal{G}(a_j^F(t), a_k^B(t))
    \label{P4.1obj}\tag{22}\\
    \st \quad & (\ref{Cons_avai_chann}), (\ref{Cons_sche}),(\ref{Cons_modu}),(\ref{Cons_SNR_Fupd}),(\ref{Cons_SNR_Bupd}). \nonumber
\end{align}
\end{small}

\noindent To handle the non-convexity of problem (P4.1), we first relax constraint (\ref{Cons_sche}) as (\ref{Cons_sche_relax}). Moreover, define a matrix $\mathbf\Psi_x(t)=\mathbf\psi_x(t)\mathbf\psi_x^H(t)$, satisfying $\mathbf\Psi_x(t)\succeq 0,\; \big[\mathbf\Psi_x(t)\big]_{n,n}=1  ,\;{\rm rank}\big(\mathbf\Psi_x(t)\big)=1$, where $x\in\{f,b\}$. Therefore, (\ref{Cons_SNR_Fupd}) and (\ref{Cons_SNR_Bupd}) can be rewritten as
\begin{small}
\begin{align}
    & {\rm Tr}\Big( \mathbf\Psi_f(t) \mathbf H_j^F(t) \Big) \ge \gamma_{th}a_j^F(t)\eta_j^F(t)\sigma_j^{F2}, \text{ and}
    \label{Cons_FSNR_tr}
    \tag{23} \\
    \chi  &{\rm Tr}\Big( \mathbf\Psi_b(t) \mathbf H_k^B(t) \Big)\sigma_o^2\gamma_{th}a_k^B(t)\eta_k^B(t) + \sigma_k^{B2}\gamma_{th}a_k^B(t)\eta_k^B(t)   \nonumber \\
    & - \chi {\rm Tr}\bigg( \mathbf\Psi_b(t)\mathbf {\Tilde H}_k^B(t) \Big( \mathbf\Psi_f(t)\Big)^T \Big( \mathbf{\Tilde  H}_k^B(t)  \Big)^H  \bigg) \le 0,
    \label{Cons_BSNR_tr}
    \tag{24}
\end{align}
\end{small}%
respectively, {\color{red}where $\mathbf H_j^F(t) = \text{diag}\{(\mathbf h_{rj}^F(t))^H\}\mathbf G(t)\mathbf w_k^B(t)(\mathbf w_k^B(t))^H\mathbf G^H(t)\text{diag}\{(\mathbf h_{rj}^F(t))\}$, $\mathbf H_k^B(t) = \text{diag}\{(\mathbf h_{rk}^B(t))^H\}\mathbf g_b(t)\mathbf g_b^H(t)\text{diag}\{(\mathbf h_{rk}^B(t))\}$, and $\tilde{\mathbf H}_k^B = \text{diag}\{(\mathbf h_{rk}^B(t))^H\}\mathbf g_b(t)(\mathbf w_k^B(t)\mathbf G(t))^T$.}

However, (\ref{Cons_BSNR_tr}) is still non-convex due to the coupling between $\mathbf\Psi_b(t)$ and $a_k^B(t)$ in the first term, and that between $\mathbf\Psi_b(t)$ and $\mathbf\Psi_f(t)$ in the third term. 

The first term and third term of (\ref{Cons_BSNR_tr}) can be transformed as the difference of convex (DC) functions (\ref{Tr_term1}) and (\ref{Tr_term3}), respectively.
\begin{small}
\begin{align}
    & \Delta_{\sigma_k}(t)  =  \chi  {\rm Tr}\Big( \mathbf\Psi_b(t) \mathbf H_k^B(t) \Big)\sigma_o^2\gamma_{th}a_k^B(t)\eta_k^B(t) \nonumber \\
    =  &{\color{red}\frac{1}{4}\chi\sigma_o^2\gamma_{th}\eta_k^B(t)} \bigg( \Big({\rm Tr}\Big( \mathbf\Psi_b(t) \mathbf H_k^B(t)\Big) + a_k^B(t) \Big)^2  \nonumber \\
    - &\Big({\rm Tr}\Big( \mathbf\Psi_b(t) \mathbf H_k^B(t)\Big) - a_k^B(t) ^2 \bigg),
    \label{Tr_term1}\tag{25} \\
    & \Delta_{sig}(t)  =  - \chi {\rm Tr}\Big( \mathbf\Psi_b(t)\mathbf {\Tilde H}_k^B(t) \big( \mathbf\Psi_f(t)\big)^T \big( \mathbf{\Tilde  H}_k^B(t)  \big)^H  \Big) \nonumber \\
    = &\frac{\chi}{2}\Big|\Big| \mathbf\Psi_b(t) - \mathbf {\Tilde H}_k^B(t) \Big( \mathbf\Psi_f(t)\Big)^T \Big( \mathbf{\Tilde  H}_k^B(t)  \Big)^H \Big|\Big|_F^2  \nonumber \\  
    - &\frac{\chi}{2}\Big|\Big| \mathbf\Psi_b(t)\Big|\Big|_F^2   -   \frac{\chi}{2}\Big|\Big| \mathbf {\Tilde H}_k^B(t) \Big( \mathbf\Psi_f(t)\Big)^T \Big( \mathbf{\Tilde  H}_k^B(t)  \Big)^H \Big|\Big|_F^2,
    \label{Tr_term3}\tag{26}
\end{align}
\end{small}%
whose upper bounds can be iteratively approximated by their first Taylor expansions, which are given as (\ref{Tr_term1_ub}) and (\ref{Tr_term3_ub}) at the bottom of this page and the top of the next page, respectively. 

To this end, (\ref{Cons_BSNR_tr}) can be approximated iteratively as
\begin{small}
\begin{equation}
    \Big[\Delta_{\sigma_k}(t)\Big]_{ub} + \Big[\Delta_{sig}(t)\Big]_{ub} + \sigma_k^{B2}\gamma_{th}a_k^B(t)\eta_k^B(t) \le 0.
    \label{Cons_BSNR_cvx} \tag{29}
\end{equation}
\end{small}

\begin{figure*}[t]
\begin{footnotesize}
\begin{align}
    & \Delta_{sig}(t) \le  \frac{\chi}{2}\Big|\Big| \mathbf\Psi_b(t) - \mathbf {\Tilde H}_k^B(t) \Big( \mathbf\Psi_f(t)\Big)^T \Big( \mathbf{\Tilde  H}_k^B(t)  \Big)^H \Big|\Big|_F^2 + \frac{\chi}{2}\Big|\Big| \mathbf\Psi_b^{(i-1)}(t)\Big|\Big|_F^2  - \chi{\rm Tr}\bigg( \Big( \mathbf\Psi_b^{(i-1)}(t) \Big)^H \mathbf\Psi_b(t)\bigg) \nonumber \\
    + & \frac{\chi}{2}\Big|\Big| \mathbf {\Tilde H}_k^B(t) \Big( \mathbf\Psi_f^{(i-1)}(t)\Big)^T \Big( \mathbf{\Tilde  H}_k^B(t)  \Big)^H \Big|\Big|_F^2 - \chi{\rm Tr}\Bigg( \bigg( \mathbf {\Tilde H}_k^B(t) \Big( \mathbf\Psi_f^{(i-1)}(t)\Big)^T \Big( \mathbf{\Tilde  H}_k^B(t)\bigg)^H \mathbf {\Tilde H}_k^B(t) \Big( \mathbf\Psi_f(t)\Big)^T \Big( \mathbf{\Tilde  H}_k^B(t)\Big)\Bigg) = \Big[\Delta_{sig}(t)\Big]_{ub}.
    \label{Tr_term3_ub}\tag{28}
    % \hline
\end{align}
\end{footnotesize}
\hrule
\end{figure*}

Moreover, the non-convex rank $1$ constraints for $\mathbf\Psi_f(t)$ and $\mathbf\Psi_b(t)$ can be equivalently substituted as the following equality constraint (taking $\mathbf\Psi_f(t)$ as an example)\cite{Rank_norm}:
\begin{equation}
    \mathcal{F}_1(\mathbf\Psi_f(t)) = ||\mathbf\Psi_f(t)||_* - ||\mathbf\Psi_f(t)||_2 = 0, \label{Cons_norm}\tag{30}
\end{equation}
where $ ||\mathbf\Psi_f(t)||_*= \sum_i\sigma_{fi}(t)$ and $||\mathbf\Psi_f(t)||_2 = \sigma_{f1}(t)$ denote the nuclear and spectrum norm of $\mathbf\Psi_f(t)$, respectively, and $\sigma_{fi}(t)$ denotes the $i^{(th)}$ largest singular value of $\mathbf\Psi_f(t)$.  Since $\mathcal{F}_1(\mathbf\Psi_f(t)) \ge 0$ always holds for any $\mathbf\Psi_f(t)$, and the equality holds if and only of $\text{rank}(\mathbf\Psi_f(t)) = 1$, penalty terms are added to the objective function as
\begin{equation}
    \mathcal{G}(a_j^F(t), a_k^B(t)) - C_f\mathcal{F}_1(\mathbf\Psi_f(t)) - C_b\mathcal{F}_2(\mathbf\Psi_b(t)), \tag{31}
\end{equation}
\noindent where $C_f$ and $C_b$ are large positive numbers, and $\mathcal{F}_2(\mathbf\Psi_b(t))$ is defined as the same manner of (\ref{Cons_norm}). However, the problem is still non-convex due to the objective function. Then we employ the SCA method to approximate $\mathcal{F}_1(\mathbf\Psi_f(t))$ as
\begin{small}
\begin{align}
    &\mathcal{F}_1\left(\mathbf\Psi_f(t)\big|\mathbf\Psi_f^{(i-1)}(t)\right) = ||\mathbf\Psi_f(t)||_* - \left\{ ||\mathbf\Psi_f^{(i-1)} ||_2 + \right. \nonumber \\
    &\left. \text{Tr}\left( \mathbf u_f^{(i-1)}(t)\left(\mathbf u_f^{(i-1)}(t)\right)^H\left( \mathbf\Psi_f(t) - \mathbf\Psi_f^{(i-1)}(t)\right) \right) \right\},
    \tag{32}
\end{align}
\end{small}
where $\mathbf u_f^{(i-1)}(t)$ denotes the largest eigenvector of $\mathbf\Psi_f^{(i-1)}(t)$.

As a result, the problem is reformulated as
\begin{small}
\begin{align}
    \mathrm{(P4.2)} \max_{a_j^F(t),a_k^B(t), \atop \mathbf \psi_f(t), \mathbf \psi_b(t)}  &\mathcal{G}(a_j^F(t), a_k^B(t)) - C_f\mathcal{F}_1\left(\mathbf\Psi_f(t)\big|\mathbf\Psi_f^{(i-1)}(t)\right) \nonumber 
    \\ & - C_b\mathcal{F}_2\left(\mathbf\Psi_b(t)\big|\mathbf\Psi_b^{(i-1)}(t)\right)
    \label{P4.2obj}\tag{33}\\
    \st \quad & (\ref{Cons_avai_chann}), (\ref{Cons_sche_relax}), (\ref{Cons_FSNR_tr}), (\ref{Cons_BSNR_cvx}), \nonumber \\
    & \mathbf \Psi_f(t)\succeq 0, \; \mathbf \Psi_b(t)\succeq 0,
    \label{Cons_semidef}\tag{\ref{P4.2obj}{a}} \\
    & \big[\mathbf\Psi_f(t)\big]_{n,n}=1, \; \big[\mathbf\Psi_b(t)\big]_{n,n}=1,
    \label{Cons_diag1}\tag{\ref{P4.2obj}{b}}
\end{align}
\end{small}%
which is a convex problem that can be solved by standard convex problem solvers.

% Then, (P4.1) can be transformed into
% % \begin{small}
% \begin{align}
%     \mathrm{(P4.2)}& \max_{a_j^F(t),a_k^B(t), \atop \mathbf \psi_f(t), \mathbf \psi_b(t)}  \mathcal{G}(a_j^F(t), a_k^B(t))
%     \label{P4.2obj}\tag{29}\\
%     \st \quad & (\ref{Cons_avai_chann}), (\ref{Cons_sche_relax}), (\ref{Cons_FSNR_tr}), (\ref{}) \nonumber \\
%     & \Big[\Delta_{\sigma_k}(t)\Big]_{ub} + \Big[\Delta_{sig}(t)\Big]_{ub} + \sigma_k^{B2}\gamma_{th}a_k^B(t)\eta_k^B(t) \le 0,
%     \label{Cons_BSNR_cvx}\tag{\ref{P4.2obj}{a}} \\
%     & \mathbf \Psi_f(t)\succeq 0, \; \mathbf \Psi_b(t)\succeq 0,
%     \label{Cons_semidef}\tag{\ref{P4.2obj}{b}} \\
%     & \mathbf \Psi_f(t)\succeq 0, \; \mathbf \Psi_b(t)\succeq 0,
%     \label{Cons_semidef}\tag{\ref{P4.2obj}{b}} \\
%     & \big[\mathbf\Psi_f(t)\big]_{n,n}=1, \; \big[\mathbf\Psi_b(t)\big]_{n,n}=1,
%     \label{Cons_diag1}\tag{\ref{P4.2obj}{c}}
% \end{align}
% % \end{small}
% \noindent which is a convex problem. However, the optimal $\mathbf \Psi_f(t)$ and $\mathbf \Psi_b(t)$ may not satisfy the rank one condition. Hence, Gaussian randomization can be used to obtain a feasible solution \cite{SDR}.

\section{Numerical Results}

In this section, we provide the numerical simulation results to show the performance of the proposed algorithm.

Assume there are $2$ front-users and $2$ back-users located outdoor and indoor, and the number of available channels is $E = 2$. The number of transmit antennas at the AP is set as $M = 4$. The distance between the outdoor AP and the RIS is set as $d_{ar} = 7 \,\rm m$. For simplicity, the distance between the RIS and each outdoor user is set to be the same as $d_{rj} = 20 \,\rm m$. The distance between the RIS and each indoor user is $d_{rk} = 3\, \rm m$. We consider all the channels follow the Rician fading model. We set the reference path loss as $-30 \text{ dB}$ at the distance $1\text{ m}$. The path loss exponents for the corresponding channels are $\alpha_{ar} = 3.5$, $\alpha_{rj} = 2.2$ and $\alpha_{rk} = 2.0$, respectively. The power of thermal noise caused by the active relay $\sigma_o^2$ and the AWGN $\sigma_j^{F2},\,\sigma_k^{B2}$ is $-80\,{\rm dBm}$. $T = 100$ time slots are simulated.
\begin{figure*}[t]
    \centering
	\subfigure[Sum AoI versus SNR threshold $\gamma_{th}$.]{
		\begin{minipage}[t]{0.3\linewidth}
			\centering	\includegraphics[width=1\linewidth]{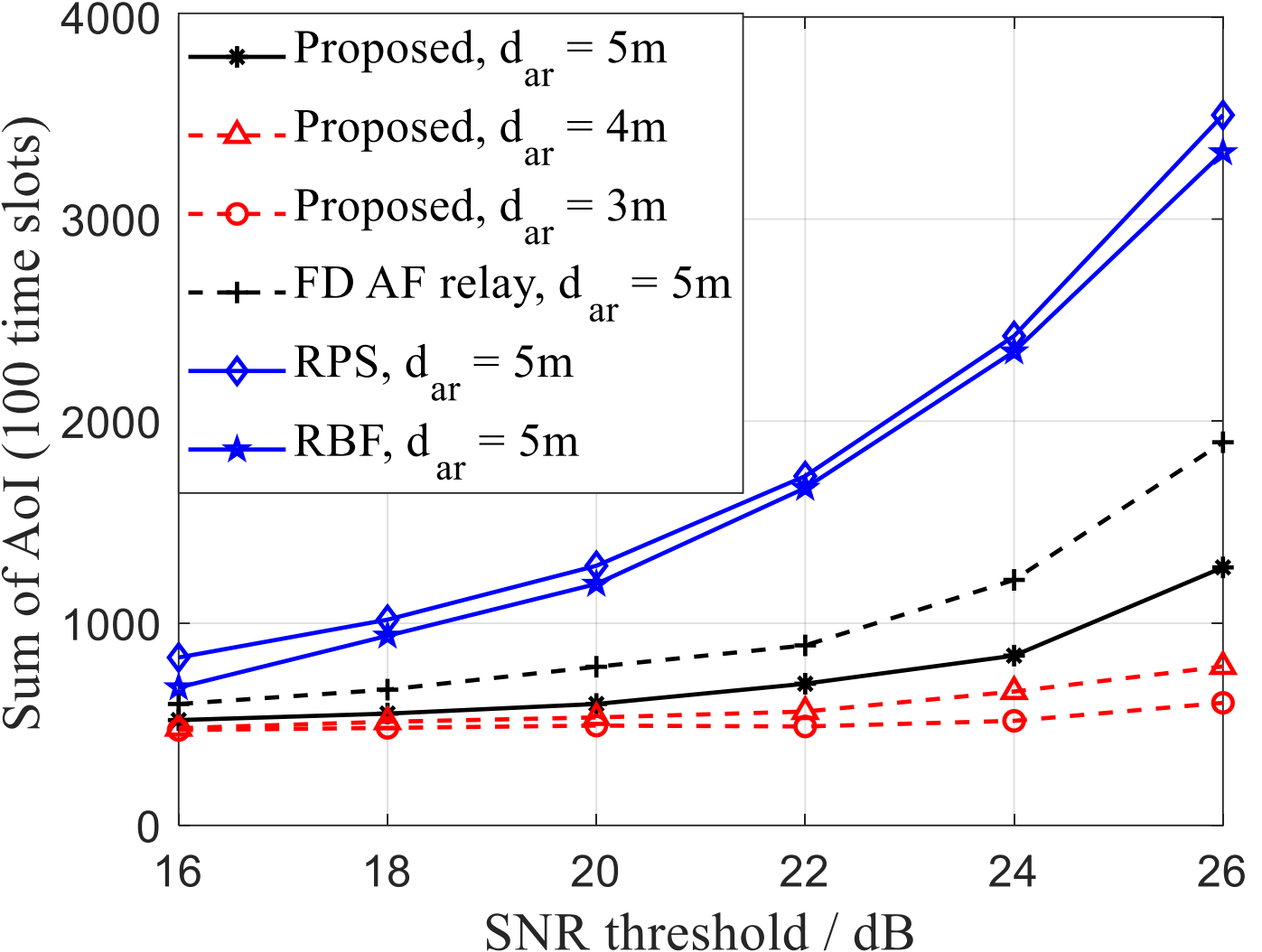}
% 			\caption{sum AoI vs SNR threshold}
			\label{AoIvsSNR}
		\end{minipage}
	}
	\subfigure[Sum AoI versus the number of reflecting elements $N_s$.]{
		\begin{minipage}[t]{0.3\linewidth}
			\centering	\includegraphics[width=1\linewidth]{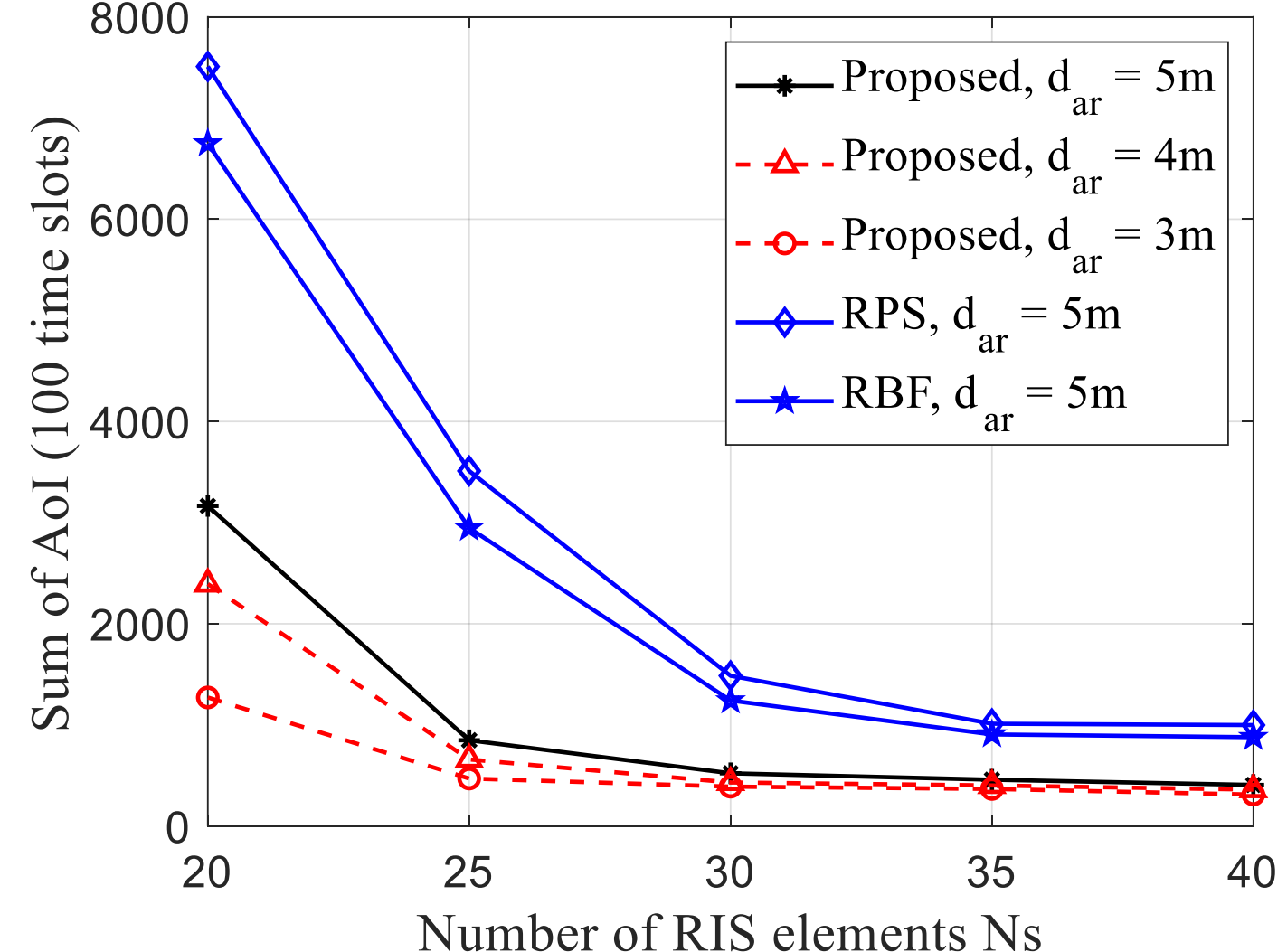}
% 			\caption{sum AoI vs number of RIS elements}
			\label{AoIvsNs}
		\end{minipage}
	}
	\subfigure[Sum AoI versus transmit power $P_0$.]{
		\begin{minipage}[t]{0.3\linewidth}
			\centering	\includegraphics[width=1\linewidth]{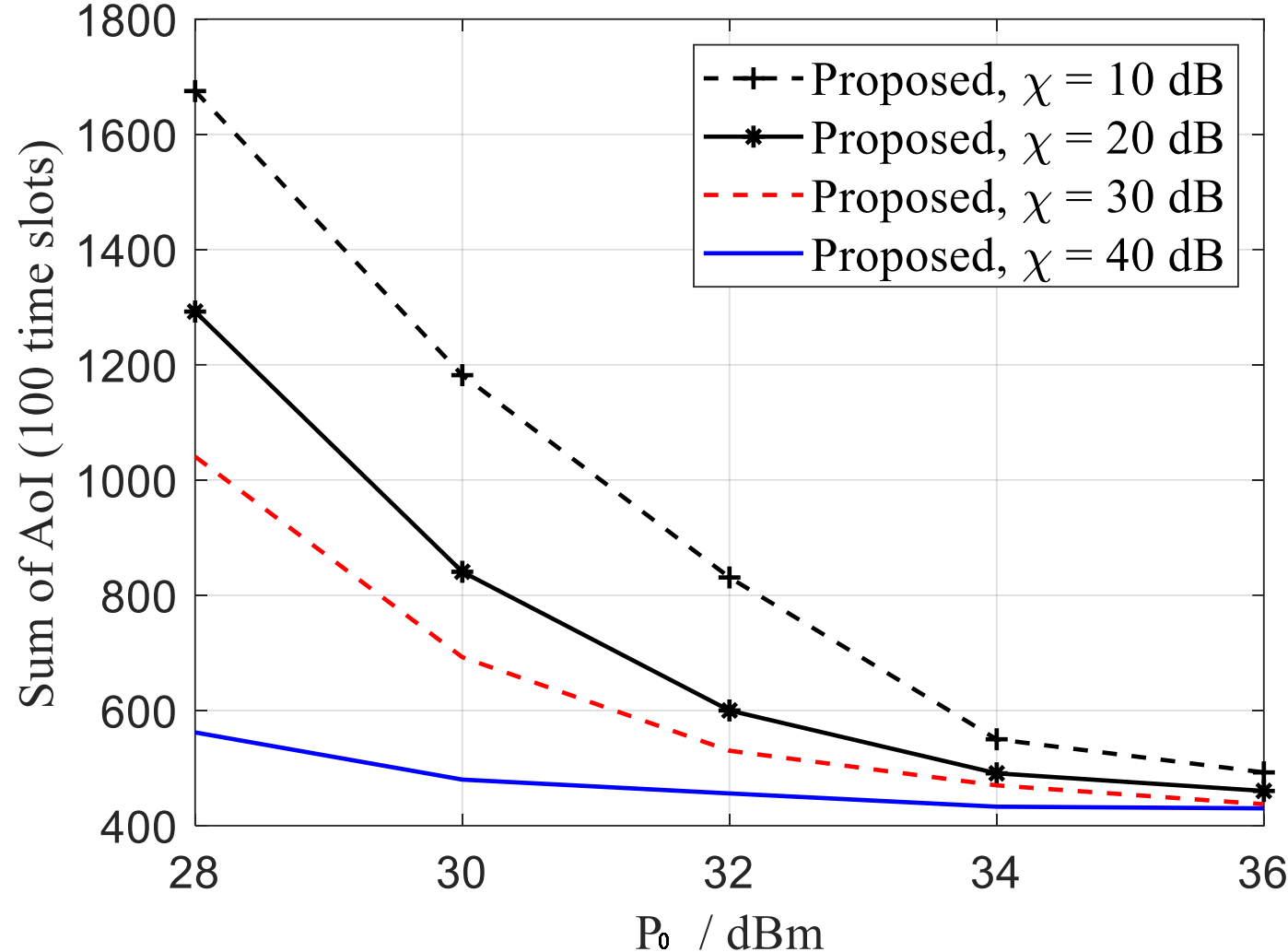}
% 			\caption{sum AoI vs Transmit Power}
			\label{AoIvsPo}
		\end{minipage}
	}
	\caption{Simulation results}
\end{figure*}

To begin with, we set the number of RIS elements $N_s = 30$ for each side, the transmit power budget $P_0 = 30\, {\rm dBm}$, and the amplifier gain of the relay $\chi = 20 \, {\rm dB}$. In Fig. \ref{AoIvsSNR}, the sum AoI versus the value of SNR threshold $\gamma_{th}$ is shown. It can be seen from Fig. \ref{AoIvsSNR} that the sum AoI in 100 time slots grows with increasing SNR threshold $\gamma_{th}$, because successful delivery is more difficult to be satisfied with a higher requirement on communication quality. The distance $d_{ar}$ between the AP and the RIS changed from $5 \rm m$ to $3 \rm m$, with shorter distance resulting in better AoI performance due to smaller path loss. Compared with the baselines i.e. {\color{red} full duplex (FD) AF relay}, random phase shifts (RPS) and random beamforming (RBF) cases at the distance $d_{ar} = 5\text{ m}$, the proposed algorithm with an AF relay achieves much better performance in terms of freshness of information.

To demonstrate the effectiveness of RIS, we fix $\gamma_{th} = 20 \text{ dB}$, and $P_0 = 30 \text{ dBm}$, and then it can be observed in Fig. \ref{AoIvsNs} that the sum AoI drops drastically with increasing number of RIS elements $N_s$, since more RIS elements provide higher diversity gain to enhance the wireless link. Also, the proposed algorithm outperforms other baselines significantly.

In Fig. \ref{AoIvsPo}, the AoI versus the transmit power $P_0$ is analyzed by setting $N_s = 20$ and $\gamma_{th} = 28\, {\rm dB}$. The sum AoI decreases with the transmit power increasing. Moreover, the effect of the relay amplifier gain $\chi$ on the freshness of information can be verified, where a larger $\chi$ improves the AoI performance with the proposed algorithm. It is because the received SNR of the indoor users can be amplified with an appropriate phase shift design.
%Compared with the full passive case ($\chi = 0\, {\rm dB}$), the proposed relay-aided RIS shows considerable AoI performance improvement particularly when $\chi$ is large.
\vspace*{-0.4cm}
\section{Conclusion}
\vspace*{-0.1cm}
In this paper, a double-sided RIS with AF relay was used to assist the outdoor and indoor multi-user wireless network. Considering timely update, the aim of the proposed algorithm was to minimize the sum AoI within a given period by optimizing the user scheduling with active and passive beamforming design. The active and passive beamforming vectors were alternatively optimized with DC functions and SCA method to deal with the non-convexity of the decomposed problems. Numerical simulation results illustrated the effectiveness of the proposed algorithm by setting appropriate parameters, compared with other baselines.
\vspace*{-0.15cm}
\bibliographystyle{IEEEtran}
\bibliography{RelayRIS}

\end{document}